\documentclass[singlecolumn,epjc3]{svjour3}
\smartqed  % flush right qed marks, e.g. at end of proof
\RequirePackage{graphicx}
\journalname{Eur. Phys. J. C}

\begin{document}

\title{A new model for spherically symmetric anisotropic compact star}

\author{S.K. Maurya\thanksref{e1,addr1}
\and Y.K. Gupta\thanksref{e2,addr2}
\and Baiju Dayanandan\thanksref{e3,addr3}
\and Saibal Ray\thanksref{e4,addr4}.}

\thankstext{e1}{e-mail: sunil@unizwa.edu.om}
\thankstext{e2}{e-mail: kumar001947@gmail}
\thankstext{e3}{e-mail: baiju@unizwa.edu.om}
\thankstext{e4}{e-mail: saibal@associates.iucaa.in}

\institute{Department of Mathematical \& Physical Sciences,
College of Arts \& Science, University of Nizwa, Nizwa, Sultanate
of Oman\label{addr1} \and Department of Mathematics, Raj Kumar
Goel Institute of Technology, Ghaziabad, U.P., India\label{addr2}
\and Department of Mathematical \& Physical Sciences, College of
Arts \& Science, University of Nizwa, Nizwa, Sultanate of
Oman\label{addr3} \and Department of Physics, Government College
of Engineering \& Ceramic Technology, Kolkata 700010, West Bengal,
India\label{addr4}}

\date{Received: date / Accepted: date}

\maketitle

\begin{abstract}
In this article we obtain a new anisotropic solution for
Einstein's field equation of embedding class one metric. The
solution is representing the realistic objects such as $Her~X-1$
and $RXJ~1856-37$. We perform detailed investigation of both
objects by solving numerically the Einstein field equations under
with anisotropic pressure. The physical features of the parameters
depend on the anisotropic factor i.e. if anisotropy is zero
everywhere inside the star then the density and pressures will
become zero and metric turns out to be flat. We report our results
and compare with the above mentioned two compact objects on a
number of key aspects: the central density, the surface density
onset and the critical scaling behavior, the effective mass and
radius ratio, the anisotropization with isotropic initial
conditions, adiabatic index and red shift. Along with this we have
also made a comparison between the classical limit and theoretical
model treatment of the compact objects. Finally we discuss the
implications of our findings for the stability condition in
relativistic compact star.
\end{abstract}


\section{Introduction}
Recent development in cosmological deep survey has clarified
progressively the origin and distribution of matter and evolution
of compact objects in the Universe. Some of their properties, such
as masses, rotation frequencies, and emission of radiation are
measurable, whereas measurements of important parameters which
decide the nature of compact stars still represent an
observational challenge. The properties that are not directly
linked to observations, such as the internal composition or masses
and radii, require the development of theoretical models. On the
theoretical side, the mass and the radius are determined by
solving the hydrostatic equilibrium equation which expresses the
equilibrium between gravitational and pressure forces. In the
framework of the general relativity the equilibrium of a spherical
object is described by the Tolman-Oppenheimer-Volkoff
(TOV)~\cite{1,2} equations and for completeness, the equation of
state is required. Very recently other theoretical advances in
modeling of densely neutral gravitating objects in strong
gravitational fields has generated much interest in last couple of
decades. This is because of its importance in describing
relativistic astrophysical objects such as neutron stars, quark
stars, hybrid proto-neutron stars, bare quark stars, etc.

The main theoretical routes have been used to study of stellar
structure and evolution is that the interior of a star can be
modeled as perfect fluid. The perfect fluid model necessarily
requires the pressure in the interior of a star to be isotropic.
Spacetime fueled by rotating anisotropic fluid has been used to
model the interior of the star. One common source is a fluid with
anisotropy in pressure. In particular at very high densities
conventional celestial bodies are not composed purely of perfect
fluids so that radial pressures are different from tangential
pressures. The model of Bowers and Liang~\cite{3} is conceptually
different form isotropic matter, but possess anisotropic matter in
the study of general relativity. Mak and Harko~\cite{4}, Sharma et
al.~\cite{5} suggest that anisotropy is a sufficient condition in
the study of dense nuclear matter with strange star. Some argument
against the existence of anisotropy, could be verifiable through
the existence of solid core or presence of type 3A fluid, were
given by Kippenhahn and Weigert~\cite{6}. On the other hand, this
can arise from different kind of phase transition and pion
condensation as pointed out by Sokolov and Sawyer~\cite{7}. The
structure of compact objects in general relativity will depend on
several parameters, including fluid and magnetic stresses, entropy
gradients, composition, heat flow and neutrino emission. However,
we restrict our attention to the case of anisotropic perfect fluid
with equilibrium composition.

The theoretical investigation of compact objects has been done by
several workers by using both analytical and numerical methods.
However, emphasize has been always given on the importance of
local pressure anisotropy. This seems to be very reasonable to
explain the matter distribution under a variety of circumstances.
Also, this has been proved to be very useful to explore
characteristics of relativistic compact
objects~\cite{8,9,10,11,12,13,14,15}. In recent years, many exact
solutions to the Einstein field equations have been generated by
different approaches~\cite{16,17,18,19,20,21,22}. Therefore, the
Einstein-Maxwell spacetime geometry for a compact object having
local anisotropic effect has attracted considerable attention in
various physical investigations. However, physical acceptability
of the solution depends on the number of criteria which include
fulfillment of various energy conditions of general relativity.

Under this background we would like to mention that in one of the
earlier works Maurya et al.~\cite{23} have proposed an algorithm
of charged anisotropic compact star while in the later
works~\cite{24,25} they have given a new approach for finding out
an anisotropic solution of Einstein's field equations by using the
metric potentials function. The present work is a sequel of the
work done by Maurya et al.~\cite{26,27}, in which the authors have
obtained the charged compact star and the structure of
relativistic electromagnetic mass model under the condition of
class one metric. It would be desirable to do a systematic
stability analysis of our model based on anisotropic spacetime. In
this work, we check the mass-radius relation, stability and
surface redshift of our models and found out their behavior is
well behaved.

The present article is organized as follows: The Sect. 2 contains
the spherical symmetric metric and the Einstein field equations.
Also we find the metric function $\lambda$ in terms of metric
function $\nu$ by applying the class one condition. In Sect. 3, we
obtain interior structure of the anisotropic star under the class
one condition. The values of arbitrary constants and total mass of
the compact star of radius $R$ are obtained by using the boundary
conditions in Sect. 4. In Sect. 5, we discuss about several
required physical conditions for anisotropic models along with the
stability analysis which is vital one. The last Sect. 6 contains
some concluding remarks on the anisotropic models.

\section{Line element for class one metric and Einstein's field equations}

We consider the static spherically symmetric metric for describing
the spacetime of compact stellar configuration as
\begin{equation}
    ds^{2} =-e^{\lambda } dr^{2} -r^{2} (d\theta ^{2} +\sin ^{2} \theta \, d\phi ^{2} )+e^{\nu } \,
    dt^{2}.
    \label{1}
\end{equation}

The energy momentum tensor of interior matter for a strange star
may be expressed in the following standard form
\begin{equation}
    \label{2}
    T_{ij}=diag(\rho, -p_{r}, -p_{t}, -p_{t}),
\end{equation}
where $\rho$,~$p_{r}$~and~$p_{t}$ are corresponding to the energy
density, radial and tangential pressures respectively of matter
distribution.

The Einstein field equations can be written as
\begin{equation}
    \label{3}
    R_{ij}-\frac{1}{2}\,R\,g_{ij}=-8\pi\,T_{ij}.
\end{equation}
Here $G = c = 1$ under the geometrized relativistic units.

In view of the metric (1), Eq. (3) yields the following
differential equations~\cite{28}
\begin{equation}
    \label{4}
    p_{r} =\frac{e^{-\lambda}}{8\pi}\left[\frac{v'}{r}  -\frac{(e^{\lambda}-1)}{r^{2}} \right],
\end{equation}

\begin{equation}
    \label{5}
    p_{t} =\frac{e^{-\lambda}}{8\pi}\left[\frac{v''}{2} -\frac{\lambda'v'}{4} +\frac{v'^{2} }{4} +\frac{v'-\lambda'}{2r} \right],
\end{equation}

\begin{equation}
    \label{6}
    \rho =\frac{e^{-\lambda}}{8\pi}\left[\frac{\lambda '}{r} +\frac{(e^{\lambda}-1 )}{r^{2}}\right].
\end{equation}

The metric (1) may represent spacetime of the embedding class one,
if it satisfies the condition of Karmarker~\cite{29}. This
condition gives the following relation between the metric
potentials $\nu$ and $\lambda$ as~\cite{24}
\begin{equation}
    \label{7}
    e^{\lambda(r)}=1+C\,\nu'^{2}\,e^{\nu}.
\end{equation}

\section{New anisotropic models for compact star}
To find interior solution of the anisotropic compact star in class
one, we consider the pressure isotropy condition as given through
the expression of the anisotropic factor as follows
\begin{equation}
    \label{8}
\frac{e^{-\lambda }}{8\pi}\left[\frac{v''}{2} -\frac{\lambda
'v'}{4} +\frac{v'^{2} }{4} -\frac{v'+\lambda '}{2r}
+\frac{e^{\lambda }-1}{r^{2} } \right] =\,(p_{t} -\,
p_{r})=\Delta.
\end{equation}

For finding out the non-zero expression for the anisotropic
factor, we assume the metric potential in the form
\begin{equation}
    \label{9}
    e^{\nu}=B\,e^{2\,Ar^{2}},
\end{equation}
where $A$ and $B$ are positive constants.

Hence from Eqs. (7) and (9), we get
\begin{equation}
    \label{10}
    e^{\lambda}=[\,1+D\,Ar^{2}\,e^{2Ar^{2}}\,],
\end{equation}
where
\begin{equation}
    \label{11}
    D=16\,A\,B\,C.
\end{equation}

\begin{figure}[!htp]\centering
    \includegraphics[width=5.5cm]{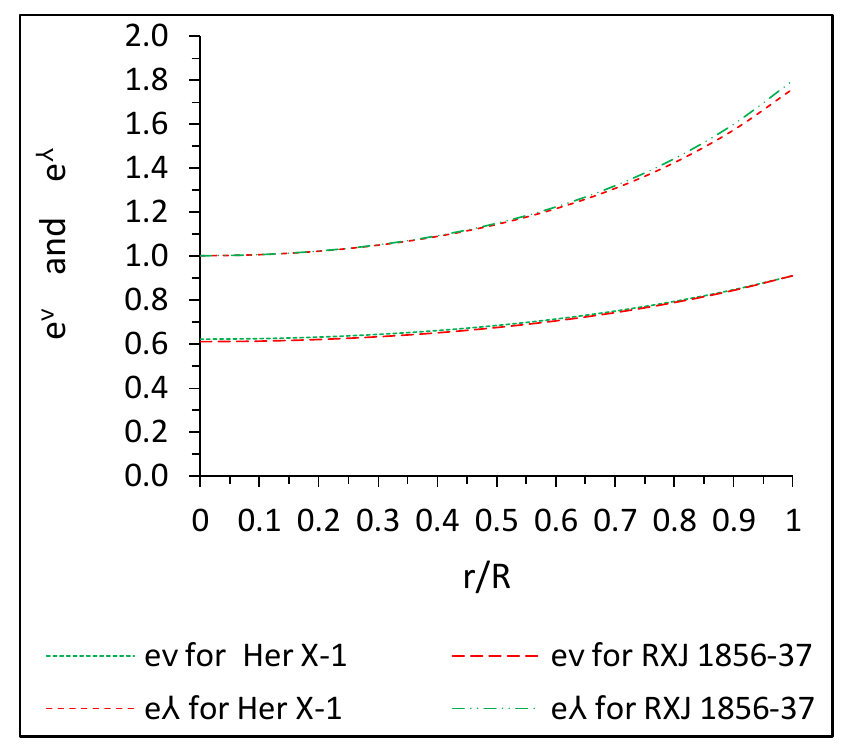}
\caption{Variation of metric potentials $e^{\nu}$ and $e^{\lambda}$ with the radial coordinate $r/R$ for $Her~X-1$ and $RXJ~1856-37$}
    \label{Fig1}
\end{figure}

From Eqs. (9) and (10), we observe that $e^{\lambda(0)}=1$  and
$e^{\nu(0)}=B$ at the centre $r=0$. This shows that metric
potentials are singularity free and positive at the centre. Also
both are monotonically increasing function which shows that these
metric potential are physically valid~\cite{30}. These features
can be observed from Fig. (1).

By plugging Eqs. (9) and (10) into Eq. (8), we get
\begin{equation}
    \label{12}
    \Delta=\,\frac{A^{2}r^{2}}{8\,\pi} \left[\,\frac{-2+D\,e^{2Ar^{2}}}{1+D\,Ar^{2}\,e^{2Ar^{2}}}\right]^2.
\end{equation}

\begin{figure}[!htp]\centering
    \includegraphics[width=5.5cm]{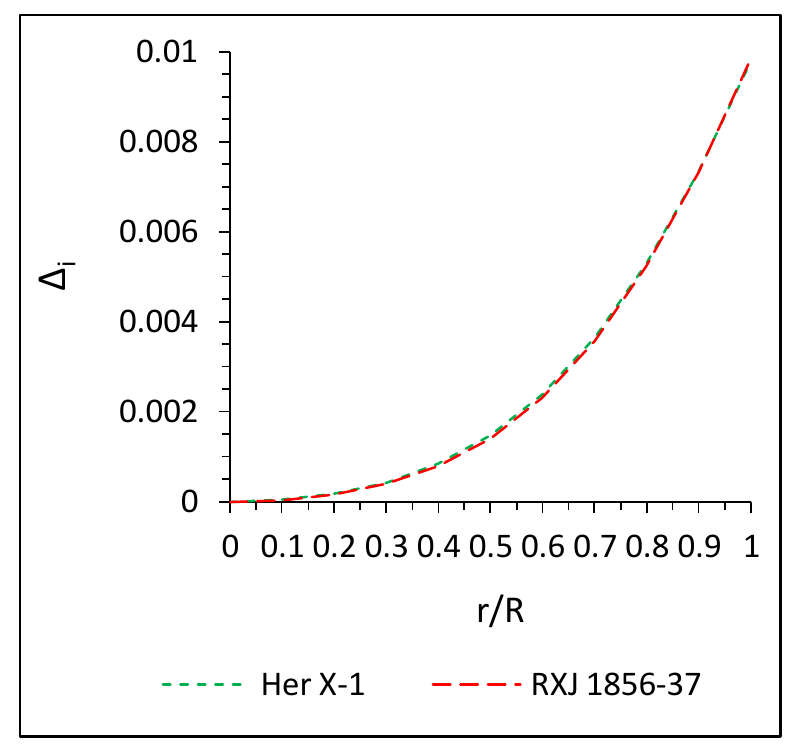}
    \caption{Variation of anisotropic factor $\Delta$ with the radial coordinate $r/R$ for $Her~X-1$ and $RXJ~1856-37$}
    \label{Fig1}
\end{figure}

We note from Fig. (2) that the anisotropy $\Delta$ is zero at the
centre $r=0$ and is monotonically increasing with increase of $r$.
Also from Eq. (12), we observe that the anisotropic factor $\Delta$
vanishes everywhere inside the compact star if and only
if $A=0$.

Eqs. (4), (5) and (6) give expressions for the radial pressure
$p_{r}$, the tangential pressure $p_{t}$ and the energy density
$\rho$ as
\begin{equation}
    \label{13}
    p_{r}=\frac{A}{8\,\pi}\,\left[\frac{4-D\,e^{2Ar^{2}}}{1+D\,Ar^{2}\,e^{2Ar^{2}}}\right],
\end{equation}

\begin{equation}
    \label{14}
    p_{t}=\frac{A}{8\,\pi}\,\left[\,\frac{4-D\,e^{2Ar^{2}}+4\,Ar^{2}}{(1+D\,Ar^{2}\,e^{2Ar^{2}})^{2}}\right],
\end{equation}

\begin{equation}
    \label{15}
    \rho=\frac{A}{8\,\pi}\,\left[\,\frac{\,D\,e^{2Ar^{2}}(3+4Ar^{2}+Ar^{2}D\,e^{2Ar^{2}})}{(1+D\,Ar^{2}\,e^{2Ar^{2}})^{2}}\right].
\end{equation}

The radial and tangential pressures at centre $r=0$ can be given
by $p_{r}=A\,(4-D)/8\,\pi$ and $p_{t}=A\,(4-D)/8\,\pi$. Since $A$
and $D$ are positive and pressure should be positive at the
centre, therefore this implies that $D<4$. In a similar way, we
can find out the density at the centre r=0 as
$\rho_{0}=(3\,A\,D/8\,\pi)$. Since the density should be positive
at the centre, then $D$ is positive due to positivity of $A$. As
$D$, $A$, $B$ all are positive, therefore this implies that $C$ is
also a positive quantity.

\begin{figure}[!htp]\centering
\includegraphics[width=5.5cm]{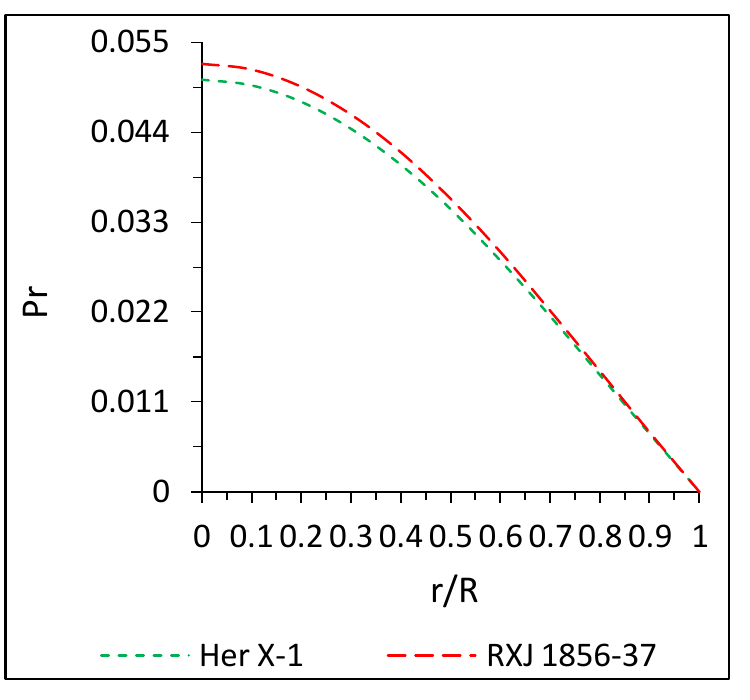}
\includegraphics[width=5.5cm]{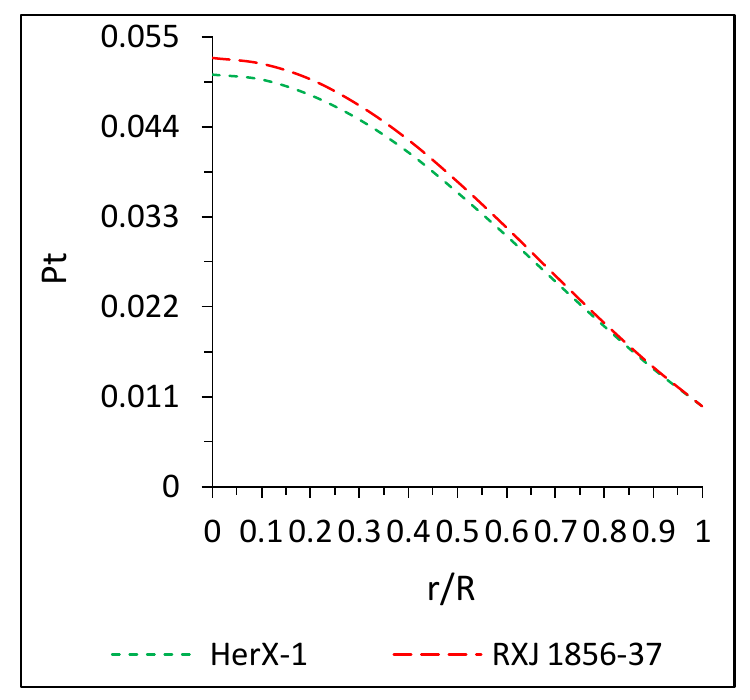}
\caption{Variation of radial pressure (left panel) and transverse
pressure (right panel) with respect to the fractional radius $r/R$
for $Her~X-1$ and $RXJ~1856-37$}
    \label{Fig3}
\end{figure}

\begin{figure}[!h]\centering
\includegraphics[width=5.5cm]{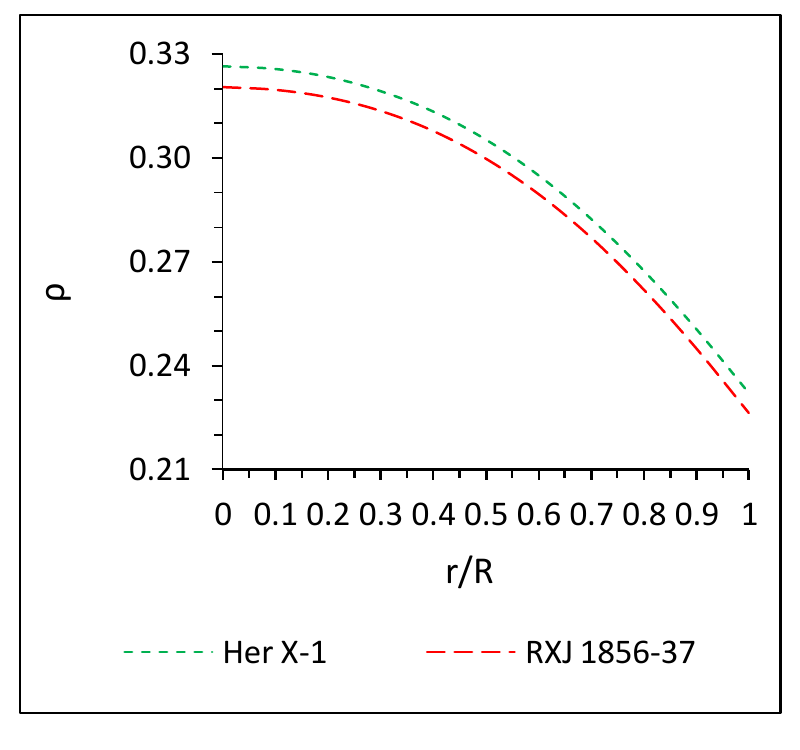}
\caption{Variation of energy density ($\rho$) with respect to the fractional radius $r/R$
for $Her~X-1$ and $RXJ~1856-37$}
    \label{Fig4}
\end{figure}

We suppose that the radial and tangential pressures of the star
are related to the matter density by a parameters $\omega_{r}$ as
$p_{r}=\omega_{r}\,\rho$ and $p_{t}=\omega_{t}\,\rho$.

Then the expressions for the parameters $\omega_r$ and $\omega_t$ are given by
\begin{equation}
    \label{16}
    \omega_{r}=\,\frac{(4-D\,e^{2Ar^{2}})(1+D\,Ar^{2}\,e^{2Ar^{2}})}{D\,e^{Ar^{2}}\,[3+4\,Ar^{2}+Ar^{2}De^{2Ar^{2}}]},
\end{equation}

\begin{equation}
    \label{17}
    \omega_{t}=\,\frac{4\,(1+Ar^{2})-D\,e^{2Ar^{2}}}{D\,e^{2Ar^{2}}\,[3+4\,Ar^{2}+Ar^{2}De^{2Ar^{2}}]}.
\end{equation}

\begin{figure}[!h]\centering
    \includegraphics[width=5.5cm]{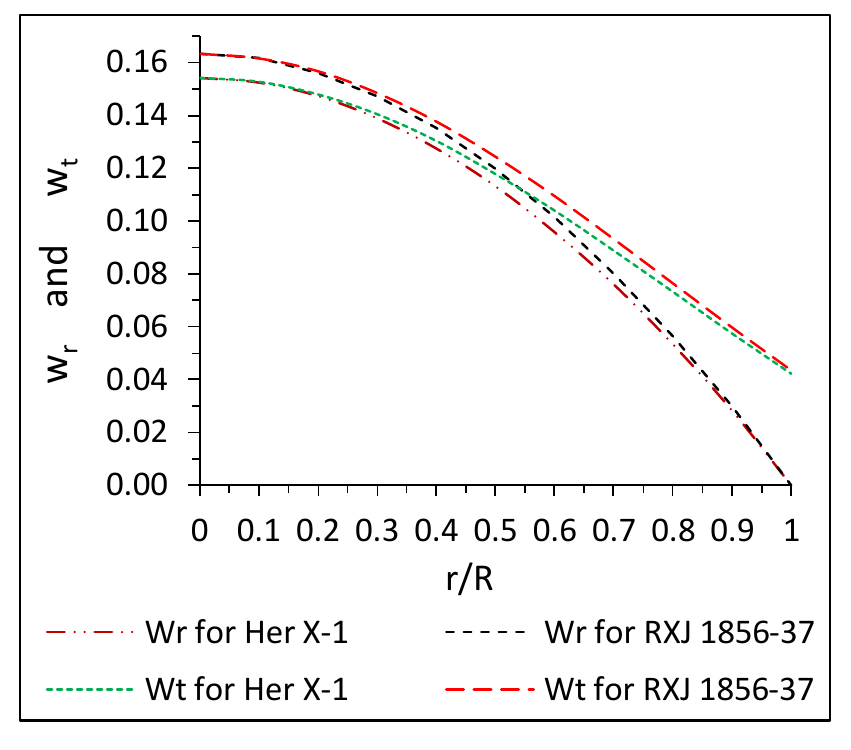}
    \caption{Variation of parameters $\omega_{r}$ and $\omega_{t}$ with the radial coordinate $r/R$ for $Her~X-1$ and $RXJ~1856-37$}
    \label{Fig.5}
\end{figure}

Form Fig. (5) it is clear that the ratio $\omega_r=p_r/\rho$ and
$\omega_t=p_t/\rho$ are less than 1. This implies that density is
dominating over the pressures throughout inside the star. However
this also implies that the underlying fluid distribution is
non-exotic in nature~\cite{39}.

\section{Matching condition}
For any physically acceptable anisotropic solution, the following
boundary conditions must be satisfied:

(i) At the surface of the compact star, the interior of metric (1)
for anisotropic matter distribution match with the exterior of
Schwarzschild solution~\cite{31}, which is given by the metric
\begin{equation}
    \label{18}
ds^{2} =\left(1-\frac{2M}{r} \right)\, dt^{2} -r^{2} (d\theta ^{2}
+\sin ^{2} \theta \, d\phi ^{2} )-\left(1-\frac{2M}{r}
\right)^{-1} dr^{2},
\end{equation}
where $M$ is a constant representing the total mass of the compact star at $r=R$.

(ii) The radial pressure $p_{r}$ must be finite and positive at
the centre $r=0$ and it must vanish at the surface $r = R$ of the
star~\cite{32}. The condition $p_r(R)=0$ gives
\begin{equation}
\label{D1}
    D=16ABC=4e^{-2AR^2}.
\end{equation}

This readily yields the radius $R$ of the compact star as
\begin{equation}
    R=\sqrt{\frac{1}{2A}\,ln\left[\frac{1}{4ABC}\right]}.
\end{equation}

Using the continuity of metric coefficients $e^{\nu}$,
$e^{\lambda}$ and $\frac{\partial g_{tt}}{\partial r}$ across the
boundary of the star gives the following equations

\begin{equation}
    \label{20}
    B\,e^{2AR^2}=1-\frac{2M}{R}
\end{equation}

\begin{equation}
    \label{21a}
    1+D\,AR^{2}\,e^{2AR^{2}}=\left[1-\frac{2M}{R}\,\right]^{-1}
\end{equation}

\begin{equation}
    \label{21b}
    4\,B\,AR^2\,e^{2AR^2}=\frac{2M}{R}
\end{equation}

These following equations with Eq. (\ref{D1}) gives the value of
unknowns M, B and C as follows:

\begin{equation}
    \label{20}
    M=\frac{R}{2}\left[\frac{4\,AR^{2}}{1+4\,AR^{2}}\right]
\end{equation}

\begin{equation}
    \label{20}
    B=\frac{e^{-2AR^2}}{1+AR^{2}}
\end{equation}

\begin{equation}
    \label{22}
    C=\frac{1+AR^{2}}{4\,A}.
\end{equation}

On the other hand, the value of the constant $A$ can be determined
by assuming the density at the surface of the star i.e. $\rho_{s}$
at $r=R$.

\section{Physical Features of the anisotropic models }

\subsection{Sound speed}
The speed of sound should monotonically decrease thought out from
the centre to the boundary of the star and it must be within the
range, i.e. $0\le V_{i}=\sqrt{dp_{i}/d\rho}<1$. It is argued by
Canuto~\cite{33} that the sound speed should decrease outwards for
the EOS with an ultra-high distribution of matter. Form Fig. (6),
it is clear that the speed of sound is monotonically decreasing
outwards.

\begin{figure}[!htp]\centering
    \includegraphics[width=5.5cm]{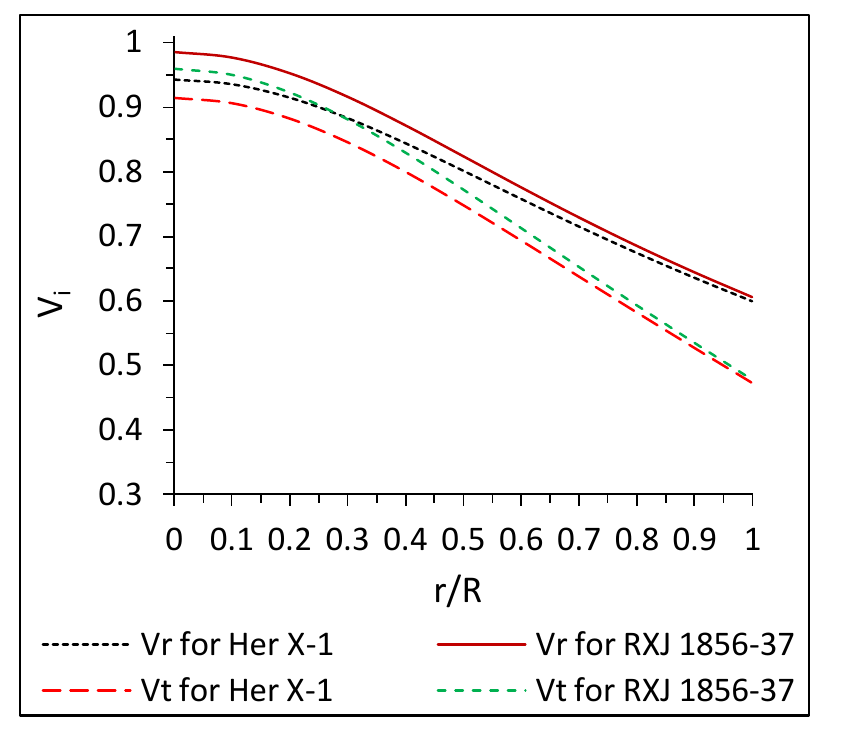}
    \caption{Variation of sound speed with the radial coordinate $r/R$ for the $Her~X-1$ and $RXJ~1856-37$}
    \label{Fig.7}
\end{figure}

\subsection{Energy conditions}
The anisotropic fluid must satisfy the following energy
conditions: Null energy condition (NEC), Weak energy condition
(WEC) and Strong energy condition (SEC). Therefore, the following
inequalities should hold simultaneously at each points inside the
compact star corresponding to the above conditions:
\begin{equation}
    \label{24}
    NEC:  \rho \ge 0,
\end{equation}

\begin{equation}
    \label{25}
    WEC_{r}: \rho -p_{r} \ge 0,
\end{equation}

\begin{equation}
    \label{26}
    WEC_{t}: \rho -p_{t} \ge 0,
\end{equation}

\begin{equation}
    \label{27}
    SEC: \rho -p_{r} - 2p_{t} \ge 0.
\end{equation}

\begin{figure}[!htp]\centering
\includegraphics[width=5.5cm]{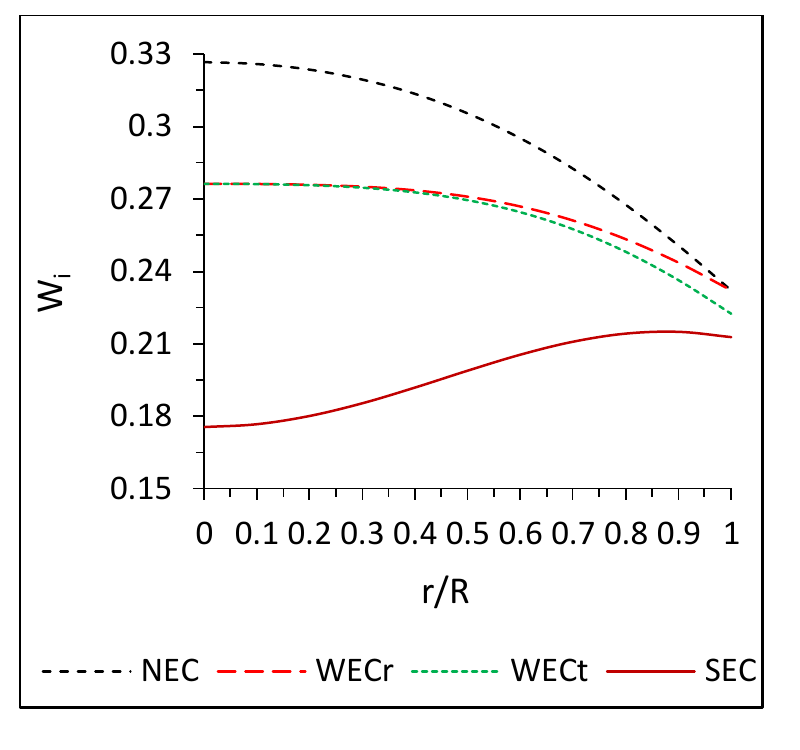}
\includegraphics[width=5.5cm]{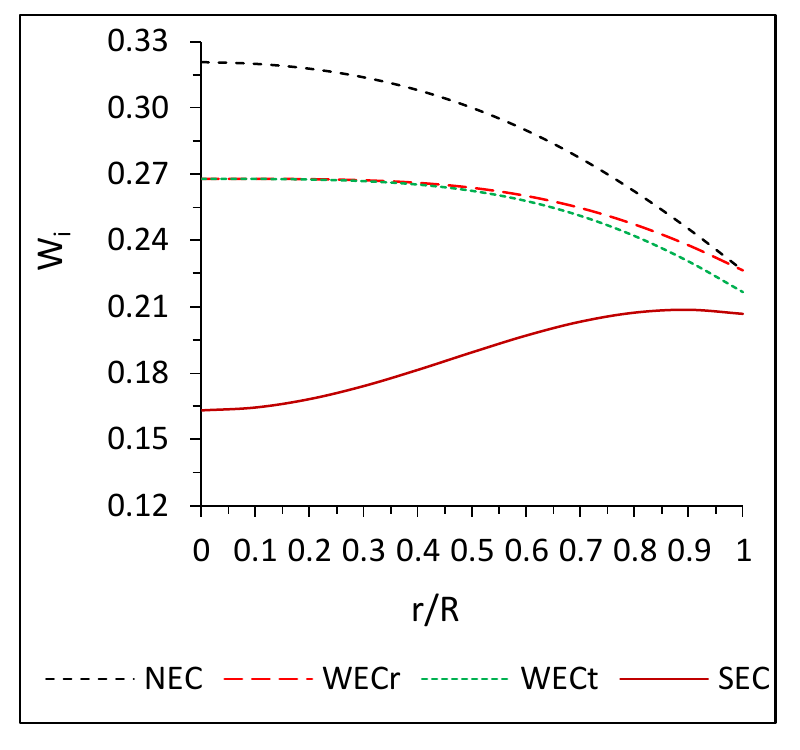}
\caption{Variation of energy conditions with the radial coordinate
$r/R$ for the $Her~X-1$ (left panel) and $RXJ~1856-37$ (right
panel)}
    \label{Fig.7}
\end{figure}

\subsection{Tolman-Oppenheimer-Volkoff equation (TOV)}
The generalized TOV equation for the anisotropic fluid distribution is given by~\cite{1,2}
\begin{equation}
    \label{28}
\frac{M_{G} (\rho +p_{r} )}{r^{2} } e^{\frac{\lambda -\nu }{2} }
+\frac{dp_{r} }{dr} +\frac{2}{r} (p_{r} -p_{t} )=0.
\end{equation}

We can write the above TOV equations as follows:
\begin{equation}
    \label{28}
    -\frac{1}{2}\,\nu'(\rho +p_{r} )-\frac{dp_{r} }{dr} +\frac{2}{r} (p_{t} -p_{r} )=0,
\end{equation}
where $M_{G}$ is the effective gravitational mass and can be given by
\begin{equation}
    \label{26}
    M_{G} (r)=\frac{1}{2} r^{2} e^{\frac{\nu -\lambda }{2} } \, \nu '.
\end{equation}

Eq. (\ref{28}) describes the equilibrium condition for an anisotropic
fluid distribution subject to the gravitational ($F_{g}$), the
hydrostatic ($F_{h}$) and the anisotropic stress ($F_{a}$) so that
\begin{equation}
    \label{29}
    F_{g} +F_{h} +F_{a} =0,
\end{equation}
where its components can be defined as
\begin{equation}
    \label{30}
    F_{g} =-\frac{1}{2} \nu '\, (\rho +p_{r}),
\end{equation}

\begin{equation}
    \label{31}
    F_{h} =-\frac{dp_{r} }{dr},
\end{equation}

\begin{equation}
    \label{32}
    F_{a} =\frac{2}{r} (p_{t} -p_{r}).
\end{equation}

The explicit form of the above forces can be expressed as
\begin{equation}
    \label{33}
    F_{g}=-\frac{A^{2}r}{4\,\pi}\,\left[\,\frac{4+2D\,e^{2Ar^{2}}\,(1+4Ar^{2})}{(1+D\,Ar^{2}\,e^{2Ar^{2}})^{2}}\right],
\end{equation}

\begin{equation}
    \label{34}
    F_{h}=-\frac{A^{2}r}{4\,\pi}\,\left[\,\frac{D\,e^{2Ar^2}(-6+2D\,e^{2Ar^{2}}-8Ar^{2})}{(1+D\,Ar^{2}\,e^{2Ar^{2}})^{2}}\right],
\end{equation}

\begin{equation}
    \label{34}
    F_{a}=\frac{A^{2}r}{4\,\pi}\,\left[\,\frac{-2+D\,e^{2Ar^{2}}}{(1+D\,Ar^{2}\,e^{2Ar^{2}})}\right]^2.
\end{equation}

\begin{figure}[!htp]\centering
\includegraphics[width=5.5cm]{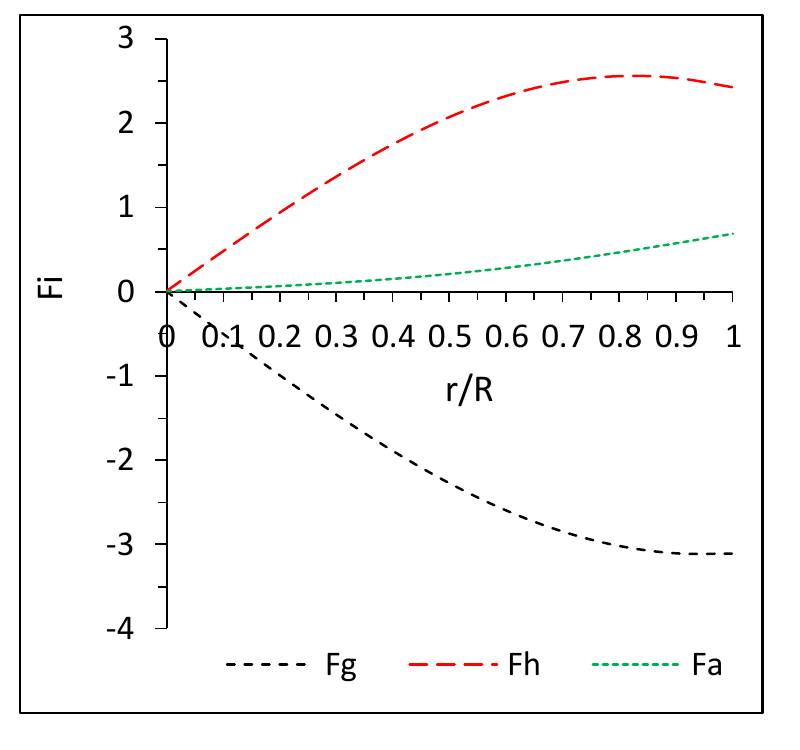}
\includegraphics[width=5.5cm]{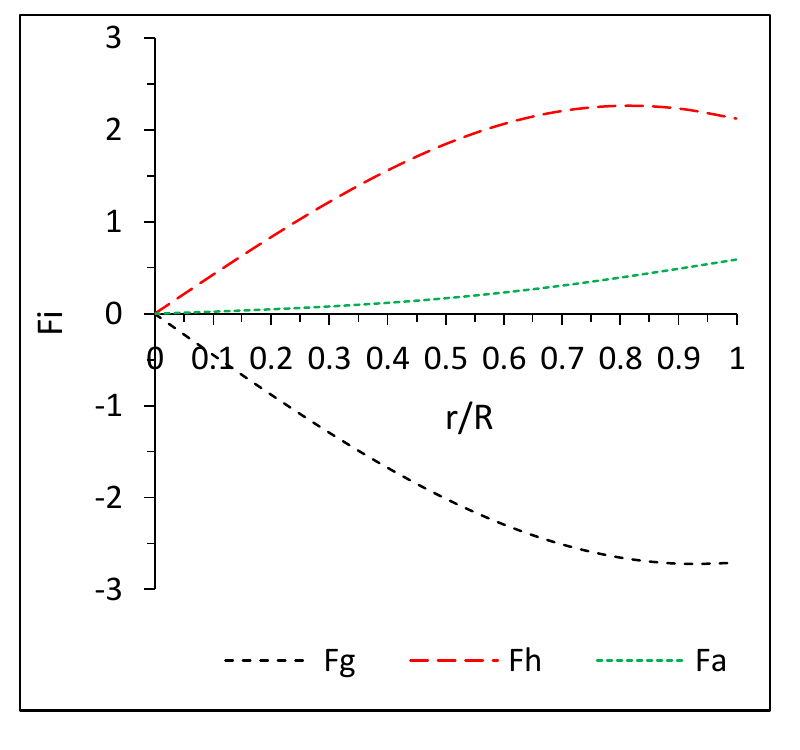}
\caption{Variation of different forces with the radial coordinate
$r/R$ for $Her~X-1$ (left panel) and $RXJ~1856-37$ (right panel)}
    \label{Fig.8}
\end{figure}

\subsection{Stability of the models}

\subsubsection{Herrera cracking concept}
We know that for physically acceptable anisotropic models, the
radial and transverse speed of sound should lies between 0 and 1
i.e. $0\le V_{r} < 1$ and $0\le V_{t} < 1$. We observe from this
inequality that parameters also should satisfy the inequality
$0\le V^{2}_{r} < 1$ and $0\le V^{2}_{t} < 1$. Now we define the
expression for the square of velocity of sound as
\begin{equation}
    \label{35}
V^{2}_{r}=\frac{dp_{r}}{d\rho}=\,\left[\frac{(6-D\,e^{2Ar^{2}}+8Ar^{2})(1+D\,Ar^{2}\,e^{2Ar^{2}})}
{D^{2}\,Ar^{2}\,e^{4Ar^{2}}-2\,(5+4Ar^{2})+D\,e^{2Ar^{2}}(5+6Ar^{2}+8A^{2}r^{4})}\right].
\end{equation}

%%%%%%%%%%%%%%%%%%%%%%%%%%%%%%%%%%%%%%%%%%%%%%%%%%%%%%%%%%%%%%%%%%%%%%%%%%%%%%%%
\begin{figure}[!htp]\centering
    \includegraphics[width=6cm]{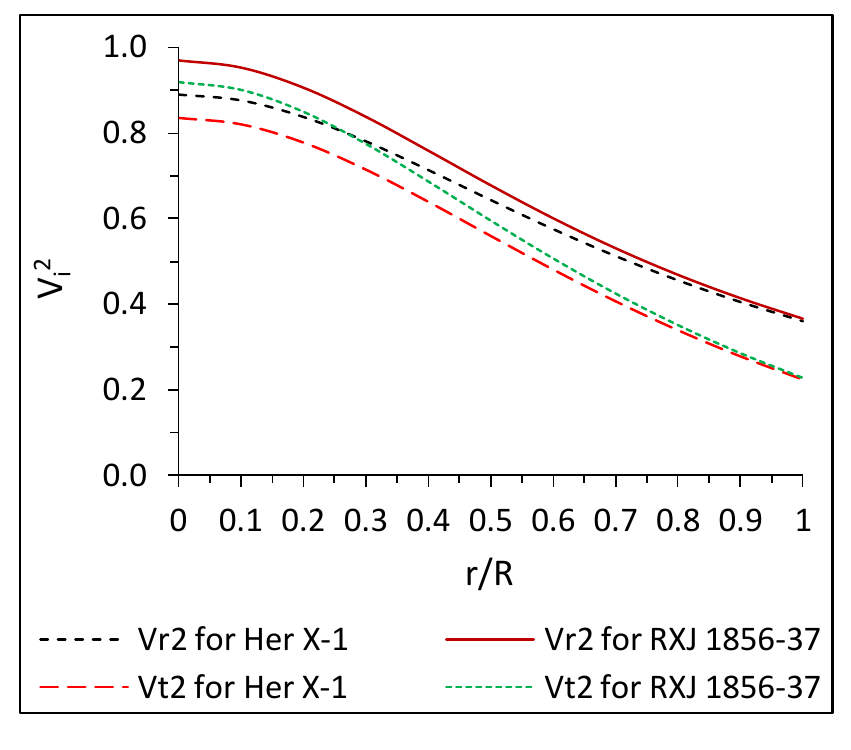}
\caption{Variation of square of radial speed of sound and
transverse speed of sound with radial coordinate $r/R$ for
$Her~X-1$ and $RXJ~1856-37$}
    \label{Fig.9}
\end{figure}
%%%%%%%%%%%%%%%%%%%%%%%%%%%%%%%%%%%%%%%%%%%%%%%%%%%%%%%%%%%%%%%%%%%%%%%%%%%%%%%%%

\begin{equation}
    \label{36}
V^{2}_{t}=\frac{dp_{t}}{d\rho}=\,\left[\frac{2[\,D\,e^{2Ar^{2}}\,(5+10Ar^{2}+8A^{2}r^{4})-2-D^{2}
\,e^{4Ar^{2}}\,(1+Ar^{2})]}{D\,e^{2Ar^{2}}[D^{2}\,Ar^{2}\,e^{4Ar^{2}}-2\,(5+4Ar^{2})+D\,e^{2AR^{2}}(5+6Ar^{2}+8A^{2}r^{4})}\right].
\end{equation}

From Fig. (9), we conclude that square of radial and transverse
speeds of sound are lies within the range everywhere inside the
stars. Therefore, $0\le |{V^{2}_{t}-V^{2}_{r}}|<1 $. In order to
examine the stability of local anisotropic fluid distribution, we
follow the cracking concept of Herrera and Aberu et
al.~\cite{34,35} which states that the region is potentially
stable where the radial speed of sound is greater than the
transverse speed of sound. This implies that there is no change in
sign $ V^{2}_{r}-V^{2}_{t}$ and $ V^{2}_{t}-V^{2}_{r}$.

So calculate the difference between the radial and transverse
speed of sound
\begin{equation}
    \label{37}
V^{2}_{t}-V^{2}_{r}=\,\left[\frac{(2-D\,e^{2Ar^{2}})[2+\,D^{2}\,Ar^{2}\,e^{4AR^{2}}-D\,e^{2Ar^{2}}
\,(1+6Ar^{2}+8A^{2}r^{4})]}{D\,e^{2Ar^{2}}[D^{2}\,Ar^{2}\,e^{4Ar^{2}}-2\,(5+4Ar^{2})+D\,e^{2Ar^{2}}(5+6Ar^{2}+8A^{2}r^{4})}\right].
\end{equation}

%%%%%%%%%%%%%%%%%%%%%%%%%%%%%%%%%%%%%%%%%%%%%%%%%%%%%%%%%%%%%%%%%%%%%%%%%%%%%%%%
\begin{figure}[!htp]\centering
    \includegraphics[width=5.5cm]{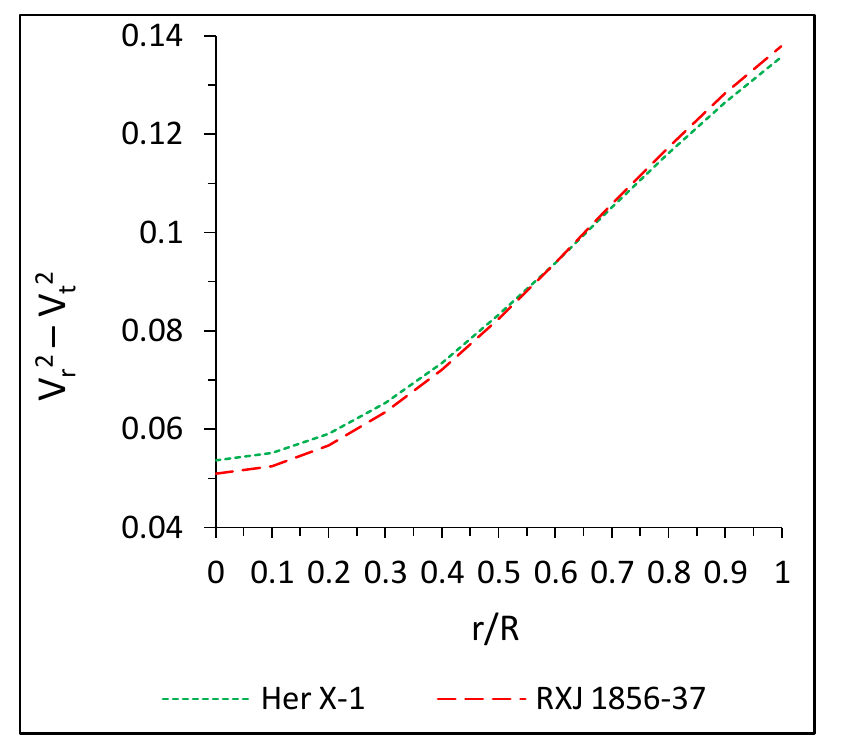}\includegraphics[width=5.5cm]{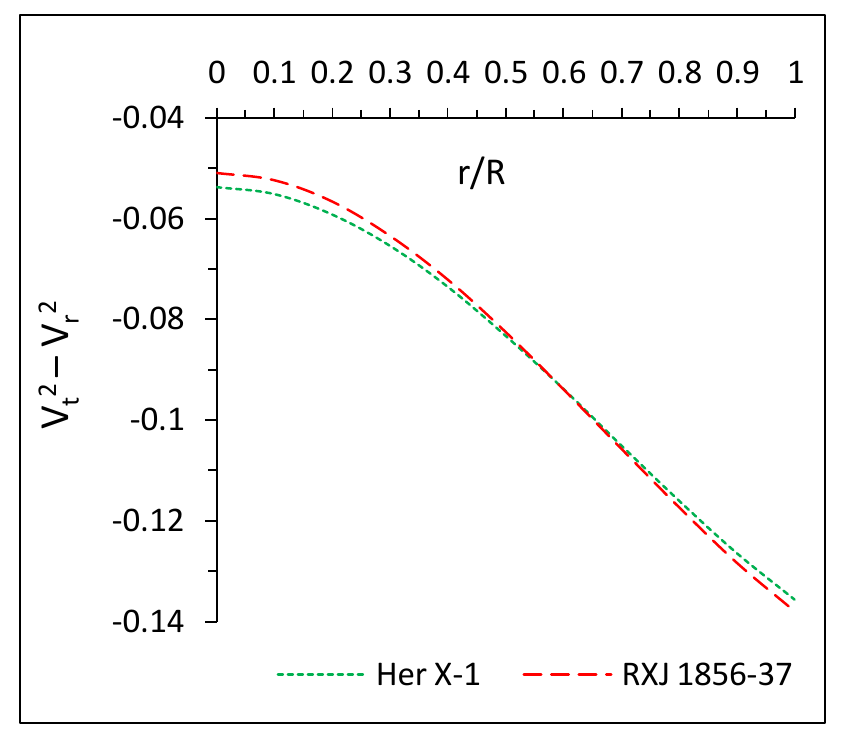}
   \caption{Variation of difference between square of radial speed
and transverse speed of sound with radial coordinate $r/R$ for $Her~X-1$ and $RXJ~1856-37$}
    \label{Fig.10}
\end{figure}
%%%%%%%%%%%%%%%%%%%%%%%%%%%%%%%%%%%%%%%%%%%%%%%%%%%%%%%%%%%%%%%%%%%%%%%%%%%%%%%%%

We note from Fig. (10) that radial speed
of sound is always greater than transverse speed of sound and also
$0\le |V^{2}_{t}-V^{2}_{r}|<1 $ everywhere inside the star. These
features represent that the proposed physical models are stable.

\subsubsection{Adiabatic index}
In order to determine an equilibrium configuration, the matter
must be stable against the collapse of local regions. This also
requires Le Chatelier's principle (known as the local or
microscopic stability condition) that the radial pressure $p_{r}$
must be a monotonically non-decreasing function of $\rho$ such
that $\frac{dp_{r}}{d\rho}>0$~\cite{36}. Heintzmann and
Hillebrandt~\cite{37} also proposed that a neutron star with an
anisotropic equation of state is stable for
$\gamma\left(=\frac{p_r+\rho}{p_r}\frac{dp_r}{d\rho}\right) >
4/3$. From Fig. (11), it is clear that the adiabatic index
($\gamma$) is more than $4/3$ everywhere inside the star.

%%%%%%%%%%%%%%%%%%%%%%%%%%%%%%%%%%%%%%%%%%%%%%%%%%%%%%%%%%%%%%%%%%%%%%%%%%%%%%%%
\begin{figure}[!htp]\centering
    \includegraphics[width=5.5cm]{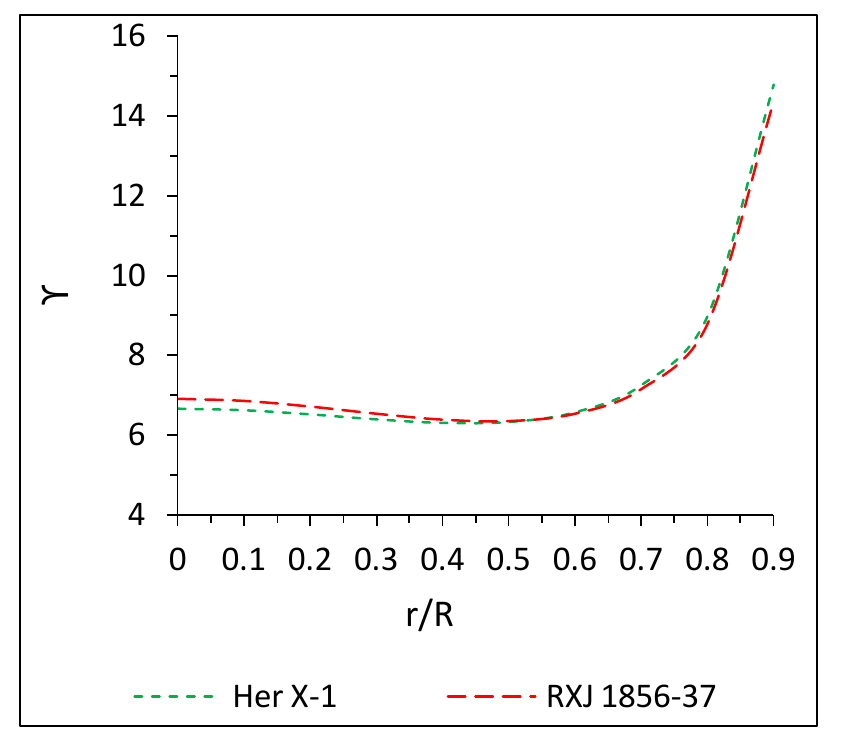}
    \caption{Variation of adiabatic index ($\gamma$) with radial coordinate r/R for $Her~X-1$ and $RXJ~1856-37$.}
    \label{Fig.11}
\end{figure}
%%%%%%%%%%%%%%%%%%%%%%%%%%%%%%%%%%%%%%%%%%%%%%%%%%%%%%%%%%%%%%%%%%%%%%%%%%%%%%%%%

\subsection{Effective mass-radius ratio}
This section contains the maximum allowable mass-radius ratio for
above proposed anisotropic fluid models. As Buchdahl~\cite{38} has
already discussed that the maximum limit of mass-radius ratio for
static spherically symmetric perfect fluid star should satisfy the
following upper bound $2M/R<8/9$. Also Mak and Harko~\cite{4} have
given the generalized expression for the same mass-radius ratio.

The effective mass of the anisotropic compact star is defined as
\begin{equation}
\label{48} M_{eff} =4\pi \int _{0}^{R}\rho \,  r^{2}
dr=\frac{1}{2} R\, [\, 1-e^{-\lambda
(R)}]=\frac{R}{2}\left[\frac{D\,AR^{2}e^{2AR^2}}{1+D\,AR^{2}e^{2AR^2}}\right]=\frac{R}{2}\left[\frac{4\,AR^{2}}{1+4\,AR^{2}}\right].
\end{equation}

However the compactness $u$ of the star can be expressed as
\begin{equation}
\label{38}
u=\frac{M_{eff}}{R}=\frac{1}{2}\,\left[\,\frac{4AR^{2}}{1+AR^{2}}\right].
\end{equation}

\subsection{Surface redshift}
The surface redshift ($Z_{s}$) corresponding to the above
compactness ($u$) is given by the expression
\begin{equation}
\label{39}
Z_{s}=\frac{1-[\,1-2\,u]^{1/2}}{[\,1-2\,u]^{1/2}}=\,\sqrt{1+4\,AR^{2}}-1.
\end{equation}

\begin{figure}[!htp]\centering
    \includegraphics[width=5.5cm]{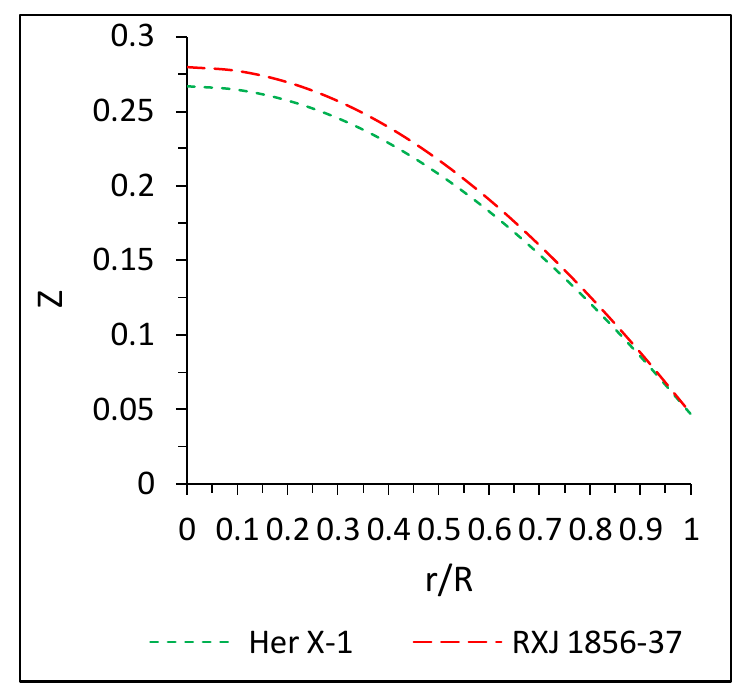}
\caption{Variation of the redshift ($Z$) with the
radial coordinate $r/R$ for $Her~X-1$ and $RXJ~1856-37$}
    \label{Fig.12}
\end{figure}

%%%%%%%%%%%%%%%%%%%%%%%%%%%%%%%%%%%%%%%%%%%%%%%%%%%%%%%%%%%%%%%%%%%%%%%%%
\begin{table}
 Table-1: Values of the model parameters Mass ($M_{\odot}$), Radius ($R$),~$D$,~$A$,~$B$ and $C$ for different compact stars\\

\begin{tabular}{@{}lrrrrrrrr@{}} \hline

Compact star  & $\frac{M}{R}$  & $M(M_{\odot})$ & $R$(km) &$D$     &$A$ &$B$ &$C$  \\
candidates & & & &  & ($km^{-2}$) &$(km^2/sec^2)$ & $(sec^2)$ \\ \hline
Her. X-1  & 0.216&  0.9825 & 6.7 & 2.7344 & 4.2371$\times 10^{-13} $  & 0.6231 & 6.4733$\times 10^{11} $ \\
RXJ 1856-37 & 0.222& 0.9041 & 6.0 & 2.6834 & 5.5443$\times 10^{-13} $ & 0.6108 & 4.9528$\times 10^{11} $ \\ \hline
\end{tabular}
\end{table}
%%%%%%%%%%%%%%%%%%%%%%%%%%%%%%%%%%%%%%%%%%%%%%%%%%%%%%%%%%%%%%%%%%%%%%%%%%%%%

%%%%%%%%%%%%%%%%%%%%%%%%%%%%%%%%%%%%%%%%%%%%%%%%%%%%%%%%%%%%%%%%%%%%%%%%%%%%%%%%
\begin{table}
Table-2: The energy densities, the central pressure and $AR^2$ for different compact star candidates for the above parameter values of Table-1\\

\begin{tabular}{@{}lrrrr@{}} \hline
Compact star & Central Density & Surface density & Central pressure & $AR^{2}$  \\
candidates & $gm/cm^{3} $ & $gm/cm^{3}$ & $dyne/cm^{2}$                           \\\hline
Her. X-1  & 1.8664$\times 10^{15} $ & 1.3273$\times 10^{15} $ & 2.5923$\times 10^{35} $ &$0.1902 $  \\ \hline
RXJ 1856-37 & 2.3968$\times 10^{15} $ & 1.6924$\times 10^{15} $ & 3.5282$\times 10^{35}$ & $0.1996$  \\ \hline
\end{tabular}
\end{table}
%%%%%%%%%%%%%%%%%%%%%%%%%%%%%%%%%%%%%%%%%%%%%%%%%%%%%%%%%%%%%%%%%%%%%%%%%%%%%%

\section{Conclusions}
In the present article, we have obtained new anisotropic compact
star models of embedding class one metric. Our models satisfy all
the physical reality conditions. Some of the special features of
the present model are as follows:

(1) We used the boundary conditions by joining the Schwarzschild
metric with class one metric at the boundary of the star $r=R$.
Subsequently we obtained the arbitrary constants $A$,\,\, $B$,\,\,
$C$ along with the total mass of the compact star and the
corresponding numerical values are provided in Table-1. All these
values are matching with the observed data of the real compact
stars.

(2) The metric potentials are free from any singularity at the
centre, positive and finite inside the star (Fig. 1). Also the
$\rho$,\,\, $p_{r}$ and $p_{t}$ are positive, finite and
monotonically decreasing away from the centre. However, the
parameters $\omega_{r}$ and $\omega_{t}$ are within the range i.e.
lies between $0$ and $1$ (Fig. 5).

(3) The model is in static equilibrium. We observe from Fig. (8)
that the gravitational force ($F_{g}$) is dominating over the
hydrostatic force ($F_{h}$) and is counter balanced by the joint
action of the hydrostatic force and the anisotropic stress.

(4) The model has density of the order $10^{15}gm/cm^{3}$. The
corresponding values for $Her~X-1$ and $RXJ~1856-37$ are as
follows:\\ (i) at the centre:
$\rho_{0}=1.8664\times{10^{15}}gm/cm^{3}$~and~$\rho_{0}=2.3968\times{10^{15}}gm/cm^{3}$,

(ii) at the surface:
$\rho_{0}=1.3273\times{10^{15}}gm/cm^{3}$~and~$\rho_{0}=1.6924\times{10^{15}}gm/cm^{3}$~(Table-2).

This density profile shows that our models may represent a
realistic anisotropic objects.

(5) The redshift is monotonically decreasing and attains its
maximum value at the centre of the compact star. The numerical
values corresponding to the $Her~X-1$ and $RXJ~1856-37$ are: (i)
at the centre: $Z_{0}= 0.2669$~and~$Z_{0}= 0.2796$, (ii) at the
surface: $Z_{s}= 0.0474$ and $Z_{s}=0.0480$.

As a final comment, an interesting and puzzling point about the
anisotropic compact model is that its stability depends on the
unavoidable anisotropic pressure and the TOV equations used to
place constraint on the anisotropic parameters. It would be
interesting to propose a richer model in which consideration of
the pressure anisotropy on the compact relativistic objects could
lead to a more realistic model of the anisotropization mechanism
as regards to the compact relativistic objects.

\section*{Acknowledgment}
The authors SKM and BD acknowledge continuous support and
encouragement from the administration of University of Nizwa. SR
wishes to thank the authorities of the Inter-University Centre for
Astronomy and Astrophysics, Pune, India for providing him Visiting
Associateship. We all are thankful to the anonymous referee for
raising several pertinent issues which have helped us to improve
the manuscript substantially.

\end{document}